\documentclass[prd, preprint, 12pt]{revtex4-1}

\usepackage{amsmath}
\usepackage{amssymb}
\usepackage{setspace}
\usepackage{graphicx}
\usepackage{bbm}
\usepackage{float}
\usepackage{hyperref}
\usepackage{braket}

\begin{document}
 
 %

\begin{center}
 {  \large {\bf Outline for a Quantum Theory of Gravity}}



{\bf Tejinder P. Singh}

{\it Tata Institute of Fundamental Research,}
{\it Homi Bhabha Road, Mumbai 400005, India}

\end{center}
\setstretch{1.24}

\vskip 0.4 in

\centerline{\bf ABSTRACT}
\medskip
\noindent By invoking an asymmetric metric tensor, and borrowing ideas from
non-commutative geometry, string theory, and trace dynamics, we propose an action function for quantum gravity. The action is proportional to the four dimensional non-commutative curvature scalar (which is torsion dependent) that is sourced by the Nambu-Goto world-sheet action for a string, plus the Kalb-Ramond string action. This `quantum gravity' is actually a non-commutative  {\it classical} matrix dynamics, and the only two fundamental constants in the theory are the square of Planck length and the speed of light. By treating the entity described by this action as a microstate, one constructs the statistical thermodynamics of a large number of such microstates, in the spirit of trace dynamics. Quantum field theory (and $\hbar$) and quantum general relativity (and $G$) emerge from the underlying matrix
dynamics in the thermodynamic limit. The statistical fluctuations that are inevitably present about equilibrium, are the source for spontaneous localisation, which drives macroscopic quantum gravitational systems to the classical general relativistic limit. While the mathematical formalism governing these ideas remains to be developed, we hope here to highlight the deep connection between quantum foundations, and the sought for quantum theory of gravity. In the sense described in this article, ongoing experimental  tests of spontaneous collapse theories are in fact also tests of string theory!



\noindent 



\bigskip


\section{Introduction and Motivation} 
The present paper should ideally be read as a sequel to \cite{stcw}, and describes how
to include gravity in the framework described in \cite{stcw}. Doing so seems to
inevitably lead to the sought for quantum theory of gravity, the basic ideas for which
are described in this article. The detailed mathematical framework governing these ideas remains to be developed. 

We have earlier argued that there ought to exist a formulation of quantum theory which does not refer to classical time. In \cite{stcw} we showed how to write down such a formulation, using an operator Minkowski space-time. The scheme of that analysis is
depicted in Fig. 1 below, which depicts three different levels of dynamics.
\begin{figure}[H]
	\centering
	\includegraphics[width=1.0\linewidth]{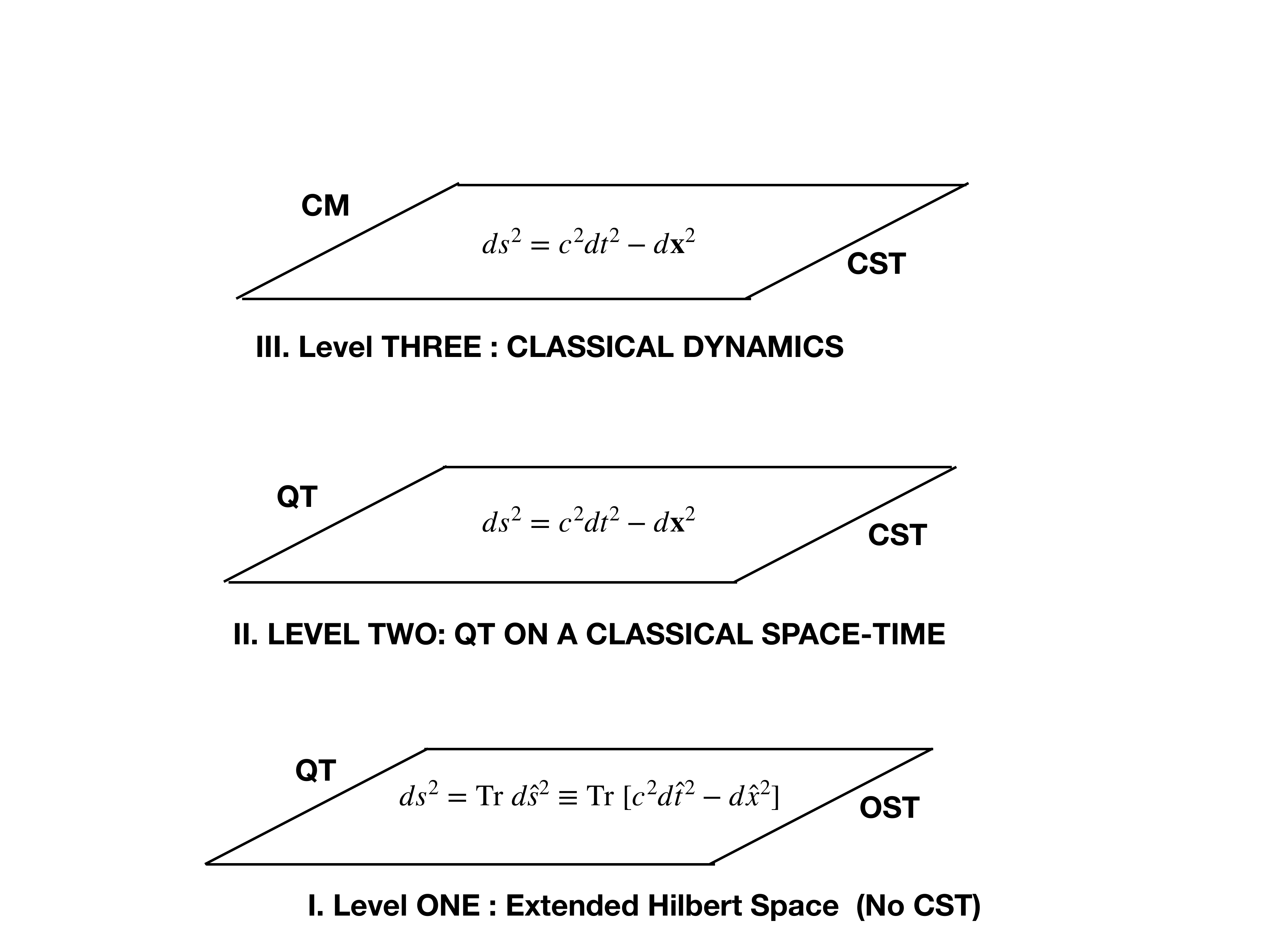}
	\caption{Introducing Level I. Quantum theory without classical space-time, and the extended Hilbert space. Here, classical space-time is replaced by the Operator Space-Time (OST), which transforms the Hilbert space of quantum theory to the Extended Hilbert Space. Taken from \cite{stcw}}.
\end{figure}
Level III is classical dynamics on a classical space-time (ignoring gravity for a moment).
Level II is quantum dynamics on a classical space-time. This must be treated as an
approximation to Level I, which is quantum theory on an operator Minkowski space-time,
as described in \cite{stcw}. The transition from Level I to Level II and Level III is made via a relativistic model of spontaneous localisation.

However, one would like to derive Level I itself from a more fundamental level, which we call Level 0. This zeroth level is introduced because we do not regard the rules of quantum theory as fundamental in themselves - it is desirable that they themselves are derived from a more basic theory with elegant symmetry principles. Such a program was initiated by Adler and collaborators, and goes by the name of trace dynamics \cite{Adler:04}. 

Trace dynamics is a classical matrix dynamics (no $\hbar$) of Grassmann matrices, further classified as Grassmann even (bosonic) and Grassmann odd (fermionic). The theory possesses a remarkable conserved charge, known as the Adler-Millard charge, and labelled $\tilde{C}$. This charge has the dimensions of action, and  is the sum (over all the bosonic degrees of freedom) of the commutators
$[q_B, p_B]$, minus the sum (over all the fermionic degrees of freedom) of the anti-commutators $\{q_F,p_F\}$. This charge, which has no analog in point particle dynamics,
plays a central role in the emergence of quantum field theory at Level II. One constructs a statistical thermodynamics of the matrix dynamics at Level 0, which results in the equipartition of the Adler-Millard charge, so that all $[q,p]$ commutators can be set to  the same constant value, which is assumed to be the Planck constant $\hbar$. Quantum field theory is shown to emerge at Level II in this thermodynamic approximation. In those situations where fluctuations around equilibrium are significant (i.e. in the macroscopic world), the mechanism of spontaneous localisation induced by these fluctuations is responsible for the quantum-to-classical transition (Level III) and the resolution of the quantum measurement problem.

However, in trace dynamics, space-time is classical, as represented by Minkowski space-time, and there is no gravity. One would like to overcome this approximation, and we took first steps in this direction by replacing Minkowski space-time by operator Minkowski space-time. This can be called generalised trace dynamics, from which we derived quantum field theory on an operator Minkowski space-time (Level I) \cite{Lochan-Singh:2011, Lochan:2012}.

There remains the challenging task of bringing in gravity, which we describe in the next
section. Since at Level 0, matter degrees of freedom are non-commutative, and matrices, one should not expect to describe gravity by the laws of classical general relativity at 
Level 0. Instead we appeal to Connes' non-commutative differential geometry, and we also argue that torsion must be included, and the metric tensor must be asymmetric. Since there is no $\hbar$ at Level 0, we expect that there is no Newton's constant $G$ either at Level 0. There are only two fundamental constants, the square of the Planck length, and the speed of light. In addition there still is the conserved Adler-Millard charge $\tilde{C}$.
The inclusion of torsion in the curvature surprisingly naturally motivates that the matter action should be that of the string in string theory, having two associated length parameters (equivalently string tensions). Much of the motivation for introducing an antisymmetric part to the metric tensor comes from the recent work of Hammond \cite{Hammond2019} and his older
review article on torsion gravity \cite{hammond2002}.

\section{An action principle for quantum gravity}

We propose to describe gravity by non-commutative curvature, as is done in Connes' non-commutative geometry \cite{Connes2000}, based on the spectral triple $({\cal A}, {\cal H}, {\cal D})$  where the symbols respectively stand for the (non-commutative) algebra of coordinates, the Hilbert space, and the Dirac operator, with distance $ds$ being defined as the inverse of the Dirac operator. We assume that the connection is asymmetric, and hence that torsion is present. Curvature in non-commutative geometry is introduced via the asymptotic spectral expansion of the Laplacian operator. As Connes asks in Section IX of his review article \cite{Connes2000} (above Eqn. 6): "What is the two-dimensional measure of a four manifold" - in other words "what is its area?". And as he notes, one has to compute the object
$\int ds^2$. As anticipated, this is proportional to the Einstein-Hilbert action, in case of Riemannian geometry:
\begin{equation}
\int ds^2 = -\frac{1}{48\pi^4}\int_{M_{4}} R\;\sqrt{g} d^4x
\end{equation}
where $R$ is the Ricci curvature scalar. We assume that this form of the action continues to hold when torsion is present, i.e. for a Riemann-Cartan space-time, and also when the
space-time metric is asymmetric. We make this action dimensionless by dividing it by the square of a length $L_p$ which we call the Planck length.  Let us denote this generalised acton as $I_G$. A related discussion of such an action (i.e. includes torsion and asymmetric metric) is available in the review article by Hammond \cite{hammond2002} (see his Eqn. 229). This will form the geometric part of our proposed action for quantum gravity. The full action, denoted by $S$, will itself be treated as dimensionless, by dividing it by a constant $\tilde{C}$ having the dimensions of action, and subsequently to be identified as the Adler-Millard charge, and eventually as the Planck constant $\hbar$, in the emergent theory at Level I.

We now come to the matter part. The curvature includes the symmetric part of the connection, for which the potential is the symmetric part $g_{\mu\nu}$ of the metric tensor. We expect the matter part to be described by fermions, and because of the presence of spin in fermions, it was natural to include torsion in the curvature, because spin is the source of torsion. Furthermore, as emphasized by Hammond, \cite{Hammond2019}, it is natural to ask for a potential for the torsion (the anti-symmetric part of the connection), and natural that this potential, labelled $\psi_{\mu\nu}$, be the
antisymmetric part of an asymmetric metric tensor. Next, if one asks for the action of
a material particle which can act as the source for Einstein equations (based on curvature including torsion, and an asymmetric metric), it can be shown that it takes the
form given in Eqn. (119) of \cite{hammond2002}.  [See the excellent analysis in
Section 3.2 of \cite{hammond2002}.] Remarkably, it is shown that this material action cannot describe a point particle, and it must have structure (hence the string can be anticipated). It is thus shown (section 8 of \cite{Hammond2019}) that a natural material action for this geometry is the two dimensional string world-sheet action made of the Nambu-Goto part, and the Kalb-Ramond part:
\begin{equation} 
I_M = \mu\int \sqrt{-\gamma} d^2\xi + \eta \int \psi_{\mu\nu} d\sigma^{\mu\nu}
\end{equation} 
where the two-metric is related to the four-metric in the standard way. The Kalb-Ramond
term provides a natural coupling between the string and the torsion potential.

Motivated by this discussion above, we propose the following action for a fundamental degree of freedom in `quantum gravity':
\begin{equation}
S/\tilde{C} = \frac{1}{L_p^2}  \int_{geom} ds^2 + \frac{1}{L^2}\int_{matter} ds^2
\end{equation}  
The matter part of the action, having a fundamental length scale $L$, can be thought of as the material contribution to the fundamental two dimensional area. Explicitly, we write the action as
\begin{equation}
S/\tilde{C} = \frac{1}{L_p^2}\int_{M_{4}} R\;\sqrt{-g} d^4x +
 \frac{1}{L_1^2}\int \sqrt{-\gamma} d^2\xi + \frac{1}{L_2^2} \int \psi_{\mu\nu} d\sigma^{\mu\nu}
\end{equation} 
assuming that there are two different length parameters $L_1$ and $L_2$ for the two
parts of the string action, each of which can be related to the string tension. Variation
of this action will yield the corresponding field equations. These are non-commutative operator Einstein equations, sourced by a string. Permissible states are eigenstates of these operator equations.

We may call this action an `atom of space-time', in support of theories which demonstrate
that gravitation as described by general relativity is an emergent thermodynamic
phenomenon \cite{paddy} coming from an underlying microscopic theory. We propose that the above action describes an atom in such an underlying theory. This action in totality describes one such atom - we should not make a distinction between the geometric part and the matter part, because there is of course no background space-time. Each atom is a non-commutatively curved string, which has its own four-dimensional Lorentzian operator `space-time', an asymmetric metric and an asymmetric connection.

If there are many strings (i.e. many atoms of space-time), each one will be described by its own action, having separate curvature and metric and connection for every one of them, and the net action will be the sum of many copies of the above action, one for each atom. 

Our so-called `quantum theory of gravity' is in fact not quantum at all. It is the classical non-commutative matrix dynamics of strings. {\it We do not quantize this theory}. It is already non-commutative - so why should it be quantized? This is the central difference of our theory from those traditional approaches to quantum gravity which quantize classical theories of gravity. Both quantum field theory and general relativity emerge from our theory as thermodynamic approximations, and in that sense, it {\it is} a quantum theory of gravity. Here one might ask as to what the status of the principle of `quantum 
linear superposition' is, in this approach? At Level 0, in general, the principle is unlikely to hold. However, at Level I, quantum theory is emergent, and in the limit that the objects under study are microscopic, non-unitary fluctuations about thermodynamic equilibrium can be neglected, and quantum linear superposition is then an excellent approximation. 
Thus, in our approach, quantum superposition is an approximate principle, rather than being an exact one.

One major challenge is to explain how to describe time evolution? One possible answer is to follow Connes, who explains in  Eqn. (5) of Section IV of \cite{Connes2000} that `non-commutative measure spaces evolve with time' and `there is a god-given group of one parameter automorphisms of the algebra $M$ of measureable coordinates. Another possibility is to exploit the trace time introduced by us in \cite{stcw}, and it might be the case that these two possibilities are related. This issue is at present under investigation.

From this point on, the analysis should proceed as in trace dynamics. What we have done is provided an explicit action principle for trace dynamics at Level 0, while also
incorporating gravity. It is plausible that  because of a global unitary invariance, the theory continues to possess the conserved Adler-Millard charge $\tilde{C}$.

Next, we construct the equilibrium statistical thermodynamics of this theory, given many such atoms. 
The philosophy is that one is not observing the matrix dynamics at Level 0, which
presumably operates at the Planck scale. Like in trace dynamics, one `derives' Planck's constant $\hbar$ from the equipartition of the conserved charge $\tilde{C}$. Once we have $\hbar$, Newton's gravitational constant $G$ is defined by $G\equiv L_p^2 c^3 /\hbar$. Quantum commutation relations emerge at thermodynamic equilibrium, and the thermal averages of the fundamental degrees of freedom are shown to obey Heisenberg equations of motion, from which one can construct conventional quantum field theory.
There is, at Level I, a Wheeler-DeWitt like equation of quantum general relativity, for the matter and metric degrees of freedom. The string length scale $L_1$ is identified with
Schwarzschild radius, and the length scale $L_2$ with Compton wavelength. 

It is to be noted though that each fundamental `atom' still retains its own curvature and thermally averaged action, both at Levels I and II. Classical space-time and the classical equations of general relativity emerge only at Level III, after relativistic spontaneous localisation from Level I, because of fluctuations around equilibrium, as described in \cite{stcw}. For reasons that are not clear to us at this stage, only fermionic macroscopic degrees of freedom undergo spontaneous localisation, leaving the curvature (metric and torsion fields) as it is. Thus, after localisation, although 
it might appear that the central part of the atom (the collapsed fermionic part) `produces' a gravitational field, in reality the field and the central part that produces it are just two aspects of the same entity, still described by the underlying action. The concept of a space-time and the gravitational field emerges only after matter localises. Furthermore, it appears as if classical space-time and its curvature are defined only by a further coarse-graining of the individual string curvatures:
\begin{equation}
R_{classical} = < R_1 + R_2 + R_3 + ... >
\end{equation}
where the quantity on the right is a coarse-graining of the curvature of the individual atoms. The quantity on the left is the classical Riemann-Cartan curvature which obeys Einstein equations with torsion included, the right hand side of Einstein equations  being the sum of the 
canonical energy momentum tensors of the individual particles.

Since there are no non-gravitational interactions in this theory, spontaneous localisation necessarily produces black holes, whose entropy can in principle be computed from the string microstates. This analysis, as well a proper mathematical development of the ideas sketched here, is currently in progress.

It is a pleasure to thank T. Padmanabhan for his insightful comments on the manuscript. I would like to thank  Kinjalk Lochan, Hendrik Ulbricht,   Bhavya Bhatt, Shounak De,  Vedant Dhruv, M Harish, Priyanka Giri,  Nancy Gupta, Navya Gupta,  Swanand Khanapurkar, Manish, Ruchira Mishra, Shlok Nahar, Branislav Nikolic, Raj Patil, Arnab Pradhan,  Kinjal Saxena, and Abhinav Varma  for helpful discussions. 

\bibliography{biblioqmtstorsion}

\end{document}